\documentclass{article}
\usepackage{booktabs}
\usepackage{CJK}
\usepackage{comment}
\usepackage{amssymb} 

\usepackage{natbib}
\usepackage[english]{babel}

\usepackage[letterpaper,top=2cm,bottom=2cm,left=3cm,right=3cm,marginparwidth=1.75cm]{geometry}

\usepackage{amsmath}
\usepackage{graphicx}
\usepackage{authblk}
\usepackage[colorlinks=true, allcolors=blue]{hyperref}

\title{The Voice Timbre Attribute Detection 2025 Challenge Evaluation Plan}

\author{
  \begin{center}
    Zhengyan Sheng$^{1}$, Jinghao He$^{1}$, Liping Chen$^{1}$, Kong Aik Lee$^{2}$, Zhen-Hua Ling$^{1}$ \\
    $^{1}$NERC-SLIP, University of Science and Technology of China, China \\
    $^{2}$Department of Electrical and Electronic Engineering, The Hong Kong Polytechnic University, Hong Kong \\
    \texttt{\{zysheng, jhhe\}@mail.ustc.edu.cn,\{lipchen, zhling\}@ustc.edu.cn} \\
    \texttt{kongaik.lee@singaporetech.edu.sg}\\
    \url{https://vtad2025-challenge.github.io/}
  \end{center}
}

\begin{document}
\maketitle


\section{Challenge Objectives}

Voice timbre refers to the unique quality or character of a person's voice that distinguishes it from others as perceived by human hearing. The Voice Timbre Attribute Detection (VtaD) 2025 challenge focuses on explaining the voice timbre attribute in a comparative manner. In this challenge, the human impression of voice timbre is verbalized with a set of sensory descriptors, including bright, coarse, soft, magnetic, and so on. The timbre is explained from the comparison between two voices in their intensity within a specific descriptor dimension. The VtaD 2025 challenge starts in May and culminates in a special proposal at the NCMMSC2025 conference in October 2025 in Zhenjiang, China.

The purpose of this VtaD challenge is to determine whether two voices exhibit obvious intensity differences within a specified descriptor dimension that conceptualizes human impression. The outcomes are expected to uncover the relationship between speech acoustics and human impression of timbre attributes. Furthermore, this task will facilitate explainable speaker recognition (\cite{DBLP:journals/corr/abs-2405-19796}), serve as an automated voice annotation tool for speaker generation (\cite{DBLP:conf/icassp/GuoLWZT23, DBLP:journals/corr/abs-2404-08857, DBLP:journals/corr/abs-2501-06394}), and promote the development of timbre-related speech technologies.

This document describes the challenge task, dataset, and baseline systems that participants can use to build their own VtaD system. Additionally, it provides detailed information on the evaluation metrics and rules, as well as guidelines for registration and submission.

\section{Task Definition}

\begin{figure}[htbp]
\centering
\includegraphics[scale=1.0]{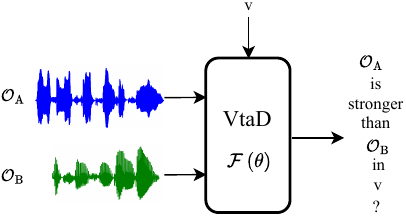}
\caption{Task definition of VtaD.}
\label{fig: task definition}
\end{figure}

As shown in Fig. \ref{fig: task definition}, given a pair of utterances ${\mathcal O}_{\rm A}$ and ${\mathcal O}_{\rm B}$ from speakers A and B, respectively, and a designated timbre descriptor v, the VtaD evaluates whether the intensity of v in ${\mathcal O}_{\rm A}$ is stronger than that in ${\mathcal O}_{\rm B}$. Mathematically, the hypothesis about the intensity difference is defined as ${\mathcal H}\left({\mathcal O}_{\rm A}, {\mathcal O}_B, {\rm v}\right)$. It means that ${\mathcal O}_{\rm B}$ is stronger than ${\mathcal O}_{\rm A}$ in the descriptor dimension v. Specifically, ${\mathcal H} \in \{0, 1\}$, where ${\mathcal H} = 1$ indicates that the hypothesis ${\mathcal H}$ is correct, and ${\mathcal H} = 0$ indicates that the hypothesis is incorrect. The hypothesis is determined by the VtaD algorithm function ${\mathcal F}({\mathcal O}_{\rm A}, {\mathcal O}_{\rm B} | {\rm v}; \theta)$, where $\theta$ is the set of algorithm parameters.

\section{Evaluation}
\subsection{Metrics}
Evaluations are conducted on speech utterances ${\mathcal O}_{\rm A}$ and $\mathcal{O}_{\rm B}$, originating from a pair of speakers A and B, respectively. The performance is evaluated in two tasks: verification and recognition. The hypothesis \({\mathcal{H}\left(\langle{\mathcal O}_{\rm A},{\mathcal O}_{\rm B}\rangle, {\rm v}\right)}=1\), where ${\rm v}\in {\mathcal V}$, is defined, assuming that ${\mathcal O}_{\rm B}$ is stronger than $\mathcal{O}_{\rm A}$ in the descriptor dimension v. The system provides the confidence score of \({\mathcal H}\) in the verification evaluation and determines whether \({\mathcal H}\) is correct in the recognition evaluation. The verification results are measured with equal error rate (EER), and the recognition results are measured with accuracy. The lower EERs and higher accuracies indicate better performance.

\begin{itemize}
\item \emph{EER}: In the verification evaluation, the target and nontarget trials are composed regarding whether the hypothesis \({\mathcal{H}\left(\langle{\mathcal O}_{\rm A},{\mathcal O}_{\rm B}\rangle, {\rm v}\right)}=1\) is true or not. Specifically, the target evaluation samples consist of instances where \({\mathcal{H}\left(\langle{\mathcal O}_{\rm A},{\mathcal O}_{\rm B}\rangle, {\rm v}\right)}=1\), while the nontarget samples comprise instances where \({\mathcal{H}\left(\langle{\mathcal O}_{\rm A},{\mathcal O}_{\rm B}\rangle, {\rm v}\right)}=0\). Given an evaluation sample $\left\{\langle{\mathcal O}_{\rm A},{\mathcal O}_{\rm B}\rangle, {\rm v}\right\}$, denote the confidence score obtained by the algorithm as $s_{\langle{\rm A},{\rm B}\rangle}^{\rm v}$. Higher $s_{\langle{\rm A},{\rm B}\rangle}^{\rm v}$ value indicates that ${\mathcal O}_{\rm B}$ is more likely to be stronger than ${\mathcal O}_{\rm A}$ in the descriptor dimension v. Finally, the EER value is computed on the confidence scores given the ground-truth target and nontarget labels of the evaluations samples.

\item \emph{Accuracy (ACC)}:
In the recognition evaluation, given the evaluation sample \(\left\{\langle{\mathcal O}_{\rm A},{\mathcal O}_{\rm B}\rangle,{\rm v}\right\}\) and the ground-truth label \(t \in \{0, 1\}\), a label of 0 indicates that the hypothesis \({\mathcal{H}\left(\langle{\mathcal O}_{\rm A},{\mathcal O}_{\rm B}\rangle, {\rm v}\right)}=1\) is false, while a label of 1 indicates that the hypothesis is true. The algorithm predicts whether the hypothesis ${\mathcal H}$ is true or not. Thereby, the accuracy is computed between the prediction and the ground truth $t$ as follows:

\begin{equation}
\text{ACC} = \frac{\text{TP} + \text{TN}}{\text{TP} + \text{TN} + \text{FP} + \text{FN}}.
\label{eq: ACC}
\end{equation}
In (\ref{eq: ACC}), TP is short for true positives, representing the number of true evaluation samples that are correctly predicted. TN is short for true negatives, representing the number of false evaluation samples that are correctly predicted. FP is short for false positives, denoting the number of false evaluation samples that are incorrectly predicted to be true. FN is short for false negative, denoting the number of true evaluation samples that are incorrectly predicted to be false.

\end{itemize}

\noindent For both EER and ACC, the results obtained by averaging across all evaluated descriptors are used as the indicators of system performance.

\subsection{Tracks}
\begin{figure}[t]
\centering
\includegraphics[scale=0.6]{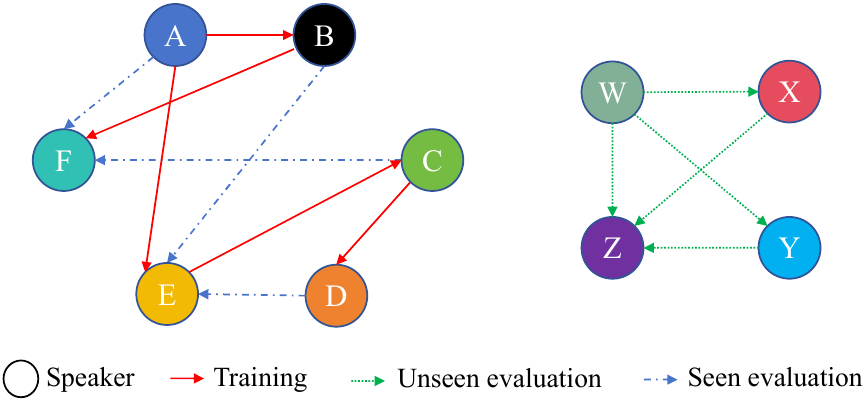}
\caption{Ordered speaker pair construction in training, unseen and seen evaluations, respectively. The direction of the arrow is from the weaker speaker to the stronger speaker in a specific descriptor in the training annotation and evaluation hypothesis.}
\label{fig: data_pairs}
\end{figure}

Regarding the speakers applied in the training and evaluation, the performance of timbre attribute intensity detection was conducted in two evaluation tracks: unseen and seen, as illustrated in Fig. \ref{fig: data_pairs}. The detailed descriptions are as follows.

\begin{itemize}
    \item \emph{Unseen}: In the unseen track, the speakers used in the evaluation phase are not present in the training phase.
    \item \emph{Seen}: In this track, the speakers employed in the evaluation phase are applied in the training phase, while distinct utterances are utilized for training and evaluation, respectively. Moreover, given a specific speaker, the ordered pairs composed with different speakers are used for training and evaluation.
\end{itemize}

\section{Data}
The VCTK-RVA dataset (\cite{vctc-rva}) is employed in our work, wherein the publicly available VCTK database was annotated for timbre intensity. In the dataset, a timbre attribute descriptor set ${\mathcal V}$ is defined, including 18 timbre descriptors, as listed in Table \ref{tab: attribute descriptor set}. In total, 101 speakers are involved, forming 6,038 annotated ordered speaker pairs \{\( \langle {\rm Speaker\,A, \,Speaker\, B} \rangle \), voice attribute v\}, indicating that Speaker B is stronger than Speaker A in the specific descriptor dimension v. The number of descriptor dimensions annotated for each ordered speaker pair ranges from 1 to 3. Notably, the speakers are annotated in a gender-dependent manner, utilizing 16 descriptors for both male and female speakers, with each gender assigned one exclusive descriptor.

In this challenge, the VCTK-RVA dataset is partitioned for training and evaluation, respectively. The evaluations are defined in two tracks regarding whether the test speakers are seen or not in the training. In the seen evaluation track, the speakers are included in the training data, while the evaluated speaker pairs are not. For each gender, the training set contains speaker pairs annotated on all 17 voice attributes. In total, 29 male and 49 female speakers are included in the training phase. In the testing phase, five descriptors are selected for each gender. Speaker statistics for the training set, seen test set, and unseen test set are presented in Table \ref{tab:the training set}, Table \ref{tab:the seen set}, and Table \ref{tab:the unseen set}, respectively. In both evaluation datasets, 20 utterances were randomly drawn from each speaker. For a speaker pair, 100 ${\mathcal H}=1$ and 300 ${\mathcal H}=0$ evaluation samples are constructed for a descriptor.

\section{Baseline Methods}
To facilitate the development of customized models by participants, we have publicly released a VtaD framework as illustrated in \cite{baselinepaper}. A comprehensive overview of the framework is provided in this section, including model architecture, parameter settings, and training procedures. The baseline performances are also presented. Hopefully, this will serve as a valuable reference for participants in designing and refining their systems, ultimately supporting more effective model optimization and advancing research in the field.

The detailed description for the baseline method can be found in []. Given the utterance pair $\mathcal{O}_{\rm A}$ and $\mathcal{O}_{\rm B}$, the speaker embedding vectors are extracted with a pre-trained speaker encoder, represented as ${\boldsymbol{e}}_{\rm A}$ and ${\boldsymbol{e}}_{\rm B}$, respectively. Given the speaker embedding vector pair, the model is trained to predict the intensity difference in each timbre descriptor. In this challenge, two speaker encoders are provided in the baseline, including a pre-trained ECAPA-TDNN encoder (\cite{DBLP:conf/interspeech/DesplanquesTD20}) and the timbre encoder in the FACodec (\cite{DBLP:conf/icml/JuWS0XYLLST000024}) architecture. The details of these two encoders are as follows:

\begin{itemize}
\item \emph{ECAPA-TDNN}: The ECAPA-TDNN speaker encoder was trained on the VoxCeleb1 (\cite{nagrani2017voxceleb}) and VoxCeleb2 (\cite{chung2018voxceleb2}) datasets, utilizing the open-source recipe ASV-Subtools\footnote{https://github.com/Snowdar/asv-subtools}.

\item \emph{FACodec}: The timbre encoder in the open-source FACodec (\cite{ju2024naturalspeech}) model\footnote{https://github.com/lifeiteng/naturalspeech3\_facodec} was used. It was trained on a 60K-hour Librilight dataset (\cite{kahn2020libri}).

\end{itemize}

\noindent The evaluation results obtained by the two models in the unseen and seen tracks are presented in Table \ref{tab: unseen eval results} and \ref{tab: seen eval results}, respectively.

\begin{table}[t]
	\centering
	\caption{Evaluation results of the VtaD model on the unseen speaker test set utilizing the ECAPA-TDNN and FACodec speaker encoders, respectively. The row \emph{Avg} is obtained by averaging the results across all the descriptors for each metric.}
	\begin{tabular}{c|ccc|ccc}

		\toprule[1pt]
		 & \multicolumn{3}{c|}{\textbf{Male}} & \multicolumn{3}{c}{\textbf{Female}} \\ 
	  \cmidrule(lr){2-4} \cmidrule(lr){5-7}
		 \textbf{Model} &Attr. & ACC (\%) & EER (\%) & Attr. & ACC (\%) & EER (\%) \\ 
		\midrule
        & Bright(\begin{CJK*}{UTF8}{gbsn}明亮\end{CJK*})                        & 64.46 & 34.05 & Bright(\begin{CJK*}{UTF8}{gbsn}明亮\end{CJK*})  & 47.24 & 49.83 \\
		& Thin(\begin{CJK*}{UTF8}{gbsn}单薄\end{CJK*})                          & 72.28 & 27.63 & Thin(\begin{CJK*}{UTF8}{gbsn}单薄\end{CJK*})    & 53.25 & 49.87 \\
	  & Low(\begin{CJK*}{UTF8}{gbsn}低沉\end{CJK*})       & 77.71 & 19.38 & Low(\begin{CJK*}{UTF8}{gbsn}低沉\end{CJK*})     & 66.63 & 35.76 \\
		\textbf{ ECAPA-TDNN}  & Magnetic(\begin{CJK*}{UTF8}{gbsn}磁性\end{CJK*})                      & 74.31 & 20.41 & Coarse(\begin{CJK*}{UTF8}{gbsn}粗\end{CJK*})  & 91.38 & 8.42 \\
		& Pure(\begin{CJK*}{UTF8}{gbsn}干净\end{CJK*})                          & 66.29 & 34.00 & Slim(\begin{CJK*}{UTF8}{gbsn}细\end{CJK*})    & 92.42 & 7.82 \\
		\cmidrule(lr){2-7}
           & Avg    & 71.01 & 27.10 & Avg   & 70.18 & 30.34  \\
        
		\midrule
        
		& Bright(\begin{CJK*}{UTF8}{gbsn}明亮\end{CJK*})      & 93.24 & 6.29  & Bright(\begin{CJK*}{UTF8}{gbsn}明亮\end{CJK*})  & 88.34 & 11.70 \\
		& Thin(\begin{CJK*}{UTF8}{gbsn}单薄\end{CJK*})        & 95.15 & 4.93  & Thin(\begin{CJK*}{UTF8}{gbsn}单薄\end{CJK*})    & 89.08 & 10.80 \\
		 & Low(\begin{CJK*}{UTF8}{gbsn}低沉\end{CJK*})         & 91.52 & 11.08 & Low(\begin{CJK*}{UTF8}{gbsn}低沉\end{CJK*})     & 87.45 & 13.27 \\
		\textbf{ FACodec} & Magnetic(\begin{CJK*}{UTF8}{gbsn}磁性\end{CJK*})    & 97.88 & 1.78  & Coarse(\begin{CJK*}{UTF8}{gbsn}粗\end{CJK*})  & 91.42 & 9.12 \\
		& Pure(\begin{CJK*}{UTF8}{gbsn}干净\end{CJK*})        & 80.54 & 19.50 & Slim(\begin{CJK*}{UTF8}{gbsn}细\end{CJK*})    & 92.54 & 6.88 \\
		\cmidrule(lr){2-7}
        & Avg    & 91.67 & 8.72  & Avg   & 89.77 & 10.35  \\

		\bottomrule[1pt]
	\end{tabular}
	\label{tab: unseen eval results}
\end{table}

\begin{table}[t]
	\centering
	\caption{Evaluation results of the VtaD model on the seen speaker test set utilizing the ECAPA-TDNN and FACodec speaker encoders, respectively. The row \emph{Avg} is obtained by averaging the results across all the descriptors for each metric.}
	\begin{tabular}{c|ccc|ccc}

		\toprule[1pt]
		 & \multicolumn{3}{c|}{\textbf{Male}} & \multicolumn{3}{c}{\textbf{Female}} \\ 
	  \cmidrule(lr){2-4} \cmidrule(lr){5-7}
	\textbf{Model} &Attr. & ACC (\%) & EER (\%) & Attr. & ACC (\%) & EER (\%) \\ 
		\midrule
        
		& Bright(\begin{CJK*}{UTF8}{gbsn}明亮\end{CJK*})                        & 94.88 & 4.78 & Bright(\begin{CJK*}{UTF8}{gbsn}明亮\end{CJK*})  & 89.15 & 9.82 \\
		& Thin(\begin{CJK*}{UTF8}{gbsn}单薄\end{CJK*})                          & 96.70 & 3.07 & Thin(\begin{CJK*}{UTF8}{gbsn}单薄\end{CJK*})    & 91.37 & 9.22 \\
	  & Low(\begin{CJK*}{UTF8}{gbsn}低沉\end{CJK*})       & 99.10 & 0.93 & Low(\begin{CJK*}{UTF8}{gbsn}低沉\end{CJK*})     & 96.35 & 3.77 \\
		\textbf{ ECAPA-TDNN}  & Magnetic(\begin{CJK*}{UTF8}{gbsn}磁性\end{CJK*})                      & 96.00 & 3.03 & Coarse(\begin{CJK*}{UTF8}{gbsn}粗\end{CJK*})  & 92.82 & 7.30 \\
		& Pure(\begin{CJK*}{UTF8}{gbsn}干净\end{CJK*})                          & 84.50 & 15.33 & Slim(\begin{CJK*}{UTF8}{gbsn}细\end{CJK*})    & 97.39 & 2.12 \\
		\cmidrule(lr){2-7}
        & Avg    & 94.23 & 5.43 & Avg   & 93.42 & 6.45 \\
        
		\midrule
        
		& Bright(\begin{CJK*}{UTF8}{gbsn}明亮\end{CJK*})      & 95.38 & 3.97  & Bright(\begin{CJK*}{UTF8}{gbsn}明亮\end{CJK*})  & 89.89 & 9.88 \\
		& Thin(\begin{CJK*}{UTF8}{gbsn}单薄\end{CJK*})        & 91.53 & 8.47  & Thin(\begin{CJK*}{UTF8}{gbsn}单薄\end{CJK*})    & 93.09 & 6.88 \\
		 & Low(\begin{CJK*}{UTF8}{gbsn}低沉\end{CJK*})         & 96.55 & 3.77 & Low(\begin{CJK*}{UTF8}{gbsn}低沉\end{CJK*})     & 98.64 & 1.45 \\
		\textbf{ FACodec} & Magnetic(\begin{CJK*}{UTF8}{gbsn}磁性\end{CJK*})    & 96.85 & 2.90  & Coarse(\begin{CJK*}{UTF8}{gbsn}粗\end{CJK*})  & 88.41 & 12.07 \\
		& Pure(\begin{CJK*}{UTF8}{gbsn}干净\end{CJK*})        & 83.03 & 15.17 & Slim(\begin{CJK*}{UTF8}{gbsn}细\end{CJK*})    & 96.81 & 2.64  \\
		\cmidrule(lr){2-7}
        & Avg    & 92.67 & 6.85  & Avg   & 93.37 & 6.58  \\

		\bottomrule[1pt]
	\end{tabular}
	\label{tab: seen eval results}
\end{table}

\section{Evaluation Rules}

%
\begin{itemize}
    \item  Participants are free to develop their own VtaD systems, using components of the baselines or not.
    \item In addition to the labeled data we provide, participants may use any other data for pre-training, unsupervised learning, and other purposes. Please clearly describe the usage of all data in the final submitted report. Note that, since VCTK is used in our evaluation, only the speakers in the provided training dataset are permitted for algorithm development.
    \item Participants are allowed to use any-pretrained models and required to describe them clearly in the submission.
    \item Participants are allowed to make 3 submissions corresponding to different models or training strategies.
\end{itemize}

\section{Registration}

Participants/teams are requested to register for the evaluation. Registration should be performed once
only for each participating entity using the  \href{https://forms.cloud.microsoft/Pages/ResponsePage.aspx?id=DQSIkWdsW0yxEjajBLZtrQAAAAAAAAAAAANAAQQ9Zp5UMjFOVldCTEMwNllQODM0VFdOR1hKSFFGMS4u}{registration form}. Participants will receive a confirmation email
within $\sim24$ hours after successful registration, otherwise or in case of any questions they should contact the organizers: \href{mailto:vtad2025_org@163.com}{vtad2025\_org@163.com}.

\section{Submission}

We will release the test set before the competition deadline. The format of the test set will consist of pairs of utterances, where participants are required to determine the difference in a specified voice attribute. Additionally, we will provide a reference document outlining the format for submitting feedback results. Participants should submit their results in the specified format before the deadline.

Each participant should also submit a single, detailed system description. All submissions should be made
according to the schedule below. Submissions received after the deadline will be marked as late submissions,
without exception.

\section{Schedule}

The result submission deadline is 4th July 2025. All participants are invited to present their work at the special proposal session for VtaD 2025 in NCMMSC2025.
\begin{table*}[h]
    \centering
    \caption{ Important dates}
    \begin{tabular}{lc}
        \toprule
        Challenge Announcement &  13th May 2025 \\ \hline 
        Release of the method description and open-source code for the baseline  & 15th May 2025 \\ \hline 
        Release of training dataset, baseline models & 17th May 2025  \\ \hline 
        Release of test set & 27th June 2025  \\ \hline
        Deadline for participants to submit system outputs & 4th July 2025 \\ \hline
        Feedback on the performance evaluation results & 11th July 2025 \\ \hline
        Final paper submission & 25th July 2025 \\ \bottomrule

    \end{tabular}
\end{table*}

\appendix
\clearpage

\section{Tables}

\begin{table}[h]
	\centering
		\caption{The descriptor (\emph{Descr.}) set used for describing the timbre. The \emph{Trans.} column gives the corresponding Chinese word. The \emph{Perc.} column presents the percentage (\%) of the anontation for each descriptor in the \emph{VCTK-RVA} dataset. The descriptors shrill and husky are exclusively annotated for female and male, respectively.}
	\begin{tabular}{lcc lcc}
		\toprule[1pt]
		
		\textbf{Descr.} & \textbf{Trans.} & \textbf{Perc.} & \textbf{Descr.} & \textbf{Trans.} & \textbf{Perc.} \\ 
		\cmidrule(lr){1-3} \cmidrule(lr){4-6}
		Bright      & \begin{CJK*}{UTF8}{gbsn}明亮\end{CJK*}      & 17.10      & Thin  &\begin{CJK*}{UTF8}{gbsn}单薄\end{CJK*}     & 13.03      \\
		Coarse &\begin{CJK*}{UTF8}{gbsn}粗\end{CJK*}      & 11.62      & Slim  &\begin{CJK*}{UTF8}{gbsn}细\end{CJK*}      & 11.31      \\
		Low &\begin{CJK*}{UTF8}{gbsn}低沉\end{CJK*}        & 7.43       & Pure &\begin{CJK*}{UTF8}{gbsn}干净\end{CJK*}       & 5.48       \\
		Rich &\begin{CJK*}{UTF8}{gbsn}厚实\end{CJK*}       & 4.71       & Magnetic  &\begin{CJK*}{UTF8}{gbsn}磁性\end{CJK*}  & 3.64       \\
		Muddy  &\begin{CJK*}{UTF8}{gbsn}浑浊\end{CJK*}     & 3.59       & Hoarse &\begin{CJK*}{UTF8}{gbsn}沙哑\end{CJK*}     & 3.32       \\
		Round  &\begin{CJK*}{UTF8}{gbsn}圆润\end{CJK*}     & 2.48       & Flat  &\begin{CJK*}{UTF8}{gbsn}平淡\end{CJK*}      & 2.15       \\
		Shrill(female only)  &\begin{CJK*}{UTF8}{gbsn}尖锐\end{CJK*}     & 2.08       & Shriveled &\begin{CJK*}{UTF8}{gbsn}干瘪\end{CJK*}  & 1.74       \\
		Muffled &\begin{CJK*}{UTF8}{gbsn}沉闷\end{CJK*}    & 1.44       & Soft    &\begin{CJK*}{UTF8}{gbsn}柔和\end{CJK*}    & 0.82       \\
		Transparent &\begin{CJK*}{UTF8}{gbsn}通透\end{CJK*} & 0.66       & Husky(male only) &\begin{CJK*}{UTF8}{gbsn}干哑\end{CJK*}      & 0.59       \\ 
		\bottomrule[1pt]
	\end{tabular}

	\label{tab: attribute descriptor set}
\end{table}

\vspace{-8pt}  

\begin{table}[h]
	\centering
		\caption{The statistics of the Male and Female speakers in the training set. The number of ordered speaker pairs (\emph{\#Pairs}) and the number of speakers (\emph{\#Speakers}) are presented for each descriptor (\emph{Descr.}).}
{\begin{tabular}{lcc lcc}
		\toprule[1pt]
        \multicolumn{3}{c}{\textbf{Male}} & \multicolumn{3}{c}{\textbf{Female}}\\
		\textbf{Descr.} & \textbf{\#Pairs} & \textbf{\#Speakers} & \textbf{Descr.} & \textbf{\#Pair} & \textbf{\#Speakers} \\ 
		\cmidrule(lr){1-3} \cmidrule(lr){4-6}
		Bright(\begin{CJK*}{UTF8}{gbsn}明亮\end{CJK*}) & 182 & 29     & Bright(\begin{CJK*}{UTF8}{gbsn}明亮\end{CJK*})  & 428 & 49 \\
        Thin(\begin{CJK*}{UTF8}{gbsn}单薄\end{CJK*}) & 82 & 29 
        & Coarse(\begin{CJK*}{UTF8}{gbsn}粗\end{CJK*}) & 382 & 49 \\
        Magnetic(\begin{CJK*}{UTF8}{gbsn}磁性\end{CJK*}) & 60 & 29 
        & Slim(\begin{CJK*}{UTF8}{gbsn}细\end{CJK*}) & 373 & 49 \\
        Low(\begin{CJK*}{UTF8}{gbsn}低沉\end{CJK*}) & 70 & 26 
        & Thin(\begin{CJK*}{UTF8}{gbsn}单薄\end{CJK*}) & 351 & 49 \\
        Pure(\begin{CJK*}{UTF8}{gbsn}干净\end{CJK*}) & 46 & 23 
        & Low(\begin{CJK*}{UTF8}{gbsn}低沉\end{CJK*}) & 191 & 48 \\
        Muffled(\begin{CJK*}{UTF8}{gbsn}沉闷\end{CJK*}) & 53 & 25 
        & Pure(\begin{CJK*}{UTF8}{gbsn}干净\end{CJK*}) & 196 & 47 \\
        Coarse(\begin{CJK*}{UTF8}{gbsn}粗\end{CJK*}) & 64 & 27 
        & Rich(\begin{CJK*}{UTF8}{gbsn}厚实\end{CJK*}) & 159 & 47 \\
        Muddy(\begin{CJK*}{UTF8}{gbsn}浑浊\end{CJK*}) & 54 & 27 
        & Hoarse(\begin{CJK*}{UTF8}{gbsn}沙哑\end{CJK*}) & 126 & 49 \\
        Slim(\begin{CJK*}{UTF8}{gbsn}细\end{CJK*}) & 56 & 27 
        & Muddy(\begin{CJK*}{UTF8}{gbsn}浑浊\end{CJK*}) & 106 & 44 \\
        Shriveled(\begin{CJK*}{UTF8}{gbsn}干瘪\end{CJK*}) & 23 & 21 
        & Shrill(\begin{CJK*}{UTF8}{gbsn}尖锐\end{CJK*}) & 69 & 45 \\
        Rich(\begin{CJK*}{UTF8}{gbsn}厚实\end{CJK*}) & 24 & 22 
        & Round(\begin{CJK*}{UTF8}{gbsn}圆润\end{CJK*}) & 35 & 31 \\ 
        Soft(\begin{CJK*}{UTF8}{gbsn}柔和\end{CJK*}) & 36 & 24 
        & Flat(\begin{CJK*}{UTF8}{gbsn}平淡\end{CJK*}) & 59 & 36 \\ 
        Hoarse(\begin{CJK*}{UTF8}{gbsn}沙哑\end{CJK*}) & 26 & 25 
        & Magnetic(\begin{CJK*}{UTF8}{gbsn}磁性\end{CJK*}) & 44 & 38 \\
        Flat(\begin{CJK*}{UTF8}{gbsn}平淡\end{CJK*}) & 30 & 23 
        & Shriveled(\begin{CJK*}{UTF8}{gbsn}干瘪\end{CJK*}) & 19 & 22 \\  
        Transparent(\begin{CJK*}{UTF8}{gbsn}通透\end{CJK*}) & 10 & 15 
        & Soft(\begin{CJK*}{UTF8}{gbsn}柔和\end{CJK*}) & 7 & 14 \\  
        Husky(\begin{CJK*}{UTF8}{gbsn}干哑\end{CJK*}) & 10 & 15 
        & Muffled(\begin{CJK*}{UTF8}{gbsn}沉闷\end{CJK*}) & 7 & 14 \\ Round(\begin{CJK*}{UTF8}{gbsn}圆润\end{CJK*}) & 14 & 14 
        & Transparent(\begin{CJK*}{UTF8}{gbsn}通透\end{CJK*}) & 2 & 4 \\     
		\bottomrule[1pt]
	\end{tabular}
        }
	\label{tab:the training set}
\end{table}

\begin{table}[t]
	\centering
		\caption{The statistics of the Male and Female speakers in the seen test set. The number of ordered speaker pairs (\emph{\#Pairs}) and the number of speakers (\emph{\#Speakers}) are presented for each descriptor (\emph{Descr.}).}
        {\begin{tabular}{lcc lcc}
		\toprule[1pt]
        \multicolumn{3}{c}{\textbf{Male}} & \multicolumn{3}{c}{\textbf{Female}}\\
		\textbf{Descr.} & \textbf{\#Pairs} & \textbf{\#Speakers} & \textbf{Descr.} & \textbf{\#Pair} & \textbf{\#Speakers} \\ 
		\cmidrule(lr){1-3} \cmidrule(lr){4-6}
		Bright(\begin{CJK*}{UTF8}{gbsn}明亮\end{CJK*}) & 20 & 21     & Bright(\begin{CJK*}{UTF8}{gbsn}明亮\end{CJK*})  & 40 & 39 \\
        Thin(\begin{CJK*}{UTF8}{gbsn}单薄\end{CJK*}) & 10 & 12 
        & Thin(\begin{CJK*}{UTF8}{gbsn}单薄\end{CJK*}) & 35 & 33  \\
        Low(\begin{CJK*}{UTF8}{gbsn}低沉\end{CJK*}) & 10 & 13 
        & Low(\begin{CJK*}{UTF8}{gbsn}低沉\end{CJK*}) & 20 & 26  \\
        Magnetic(\begin{CJK*}{UTF8}{gbsn}磁性\end{CJK*}) & 10 & 13 
        & Coarse(\begin{CJK*}{UTF8}{gbsn}粗\end{CJK*}) & 20 & 26  \\
        Pure(\begin{CJK*}{UTF8}{gbsn}干净\end{CJK*}) & 10 & 13 
        & Slim(\begin{CJK*}{UTF8}{gbsn}细\end{CJK*}) & 20 & 26  \\
		\bottomrule[1pt]
	\end{tabular}
        }
	\label{tab:the seen set}
\end{table}

\begin{table}[t]
	\centering
		\caption{The statistics of the Male and Female speakers in the unseen test set. The number of ordered speaker pairs (\emph{\#Pairs}) and the number of speakers (\emph{\#Speakers}) are presented for each descriptor (\emph{Descr.}).}
 {\begin{tabular}{lcc lcc}
		\toprule[1pt]
        \multicolumn{3}{c}{\textbf{Male}} & \multicolumn{3}{c}{\textbf{Female}}\\
		\textbf{Descr.} & \textbf{\#Pairs} & \textbf{\#Speakers} & \textbf{Descr.} & \textbf{\#Pairs} & \textbf{\#Speakers} \\ 
		\cmidrule(lr){1-3} \cmidrule(lr){4-6}
		Bright(\begin{CJK*}{UTF8}{gbsn}明亮\end{CJK*}) & 34 & 20     & Bright(\begin{CJK*}{UTF8}{gbsn}明亮\end{CJK*})  & 35 & 40 \\
        Thin(\begin{CJK*}{UTF8}{gbsn}单薄\end{CJK*}) & 29 & 10 
        & Thin(\begin{CJK*}{UTF8}{gbsn}单薄\end{CJK*}) & 28 & 35  \\
        Low(\begin{CJK*}{UTF8}{gbsn}低沉\end{CJK*}) & 13 & 10 
        & Low(\begin{CJK*}{UTF8}{gbsn}低沉\end{CJK*}) & 15 & 10  \\
        Magnetic(\begin{CJK*}{UTF8}{gbsn}磁性\end{CJK*}) & 17 & 10 
        & Coarse(\begin{CJK*}{UTF8}{gbsn}粗\end{CJK*}) & 26 & 40  \\
        Pure(\begin{CJK*}{UTF8}{gbsn}干净\end{CJK*}) & 6 & 5 
        & Slim(\begin{CJK*}{UTF8}{gbsn}细\end{CJK*}) & 26 & 40  \\
		\bottomrule[1pt]
	\end{tabular}
        }
	\label{tab:the unseen set}
\end{table}

\clearpage

\bibliography{sample} 
\end{document}